# Observation of Persistent Zero Modes and Superconducting Vortex Doublets in UTe$_2$

Nileema Sharma[1,2]†, Matthew Toole[1,2]†, James McKenzie[1,2], Fangjun Cheng[1,2], Mitchell M. Bordelon[3], Sean M. Thomas[3], Priscila F. S. Rosa[3], Yi-Ting Hsu[1]*, and Xiaolong Liu[1,2]*

[1]*Department of Physics and Astronomy, University of Notre Dame, Notre Dame, IN 46556, USA*

[2]*Stavropoulos Center for Complex Quantum Matter, University of Notre Dame, Notre Dame, IN 46556, USA*

[3]*Los Alamos National Laboratory, Los Alamos, New Mexico 87545, USA*

\* Corresponding authors. Email: yhsu2@nd.edu, xliu33@nd.edu

† These authors contributed equally to this work.

**Abstract**

Superconducting vortices can reveal electron pairing details and nucleate topologically protected states[1]. Yet, vortices of bulk spin-triplet superconductors have never been visualized at the atomic scale. Recently, UTe$_2$ has emerged as a prime spin-triplet superconductor[2,3,4,5,6,7,8], but its superconducting order parameter is elusive, and whether time-reversal symmetry (TRS) is broken remains unsettled[4,5,9,10]. Here, we visualize vortices on the (011) surface of ultra-clean UTe$_2$ single crystals[11,12] ($T_c$ = 2.1 K) using scanning tunneling microscopy (STM). We introduce $\frac{d^2I}{dV^2}$ imaging as an effective technique for vortex visualization in superconductors with substantial residual zero-energy density of states (DOS), as in UTe$_2$. Anisotropic single-flux-quantum vortices, with coherence lengths of ~12 nm (4 nm) parallel (perpendicular) to the **a**-axis, form a triangular vortex lattice (VL) under a small out-of-plane magnetic field. The invariance of vortex structures and VL under changes of field polarity and cooling history strongly supports time-reversal invariant superconductivity under zero field. At vortex cores (VCs), non-split, spectrally sharp zero-bias conductance peaks (ZBPs) persist up 8 T that are consistent with symmetry-protected Majorana zero modes (MZMs) in a topological vortex line. Close examination of vortex structures reveals a mirror-asymmetric doublet—one with ZBPs and another with an enhanced superconducting gap, possibly originating from a field-induced multi-component order parameter.

**Main**

Topological superconductivity enables fault-tolerant quantum computation by hosting topologically protected non-Abelian anyons, known as Majorana modes[13,14,15], which emerge at the boundaries and defects of odd-parity superconductors with spin-triplet pairing[16,17]. Compared to proximitized topological insulators[18,19,20,21], semiconductors[17], or magnetic materials[22,23,24,25], a bulk Cooper-pair condensate with intrinsic spin-triplet pairing is extremely rare. The superconducting order parameter $\Delta$ of a spin-triplet condensate is represented using the **d** vector, $\Delta(\mathbf{k}) \equiv (\mathbf{d} \cdot \boldsymbol{\sigma}) i\sigma_2$, such that

$$\Delta = \begin{pmatrix} \Delta_{\uparrow\uparrow} & \Delta_{\uparrow\downarrow} \\ \Delta_{\downarrow\uparrow} & \Delta_{\downarrow\downarrow} \end{pmatrix} = \begin{pmatrix} -d_x + id_y & d_z \\ d_z & d_x + id_y \end{pmatrix} \quad (1)$$

where $\sigma_i$ are the Pauli matrices. When vortices form in a spin-triplet condensate, the degrees of freedom from finite orbital ($L = 1$) and spin angular momenta ($S = 1$) can give rise to internal vortex structures and quantum phase winding that depart from that of conventional Abrikosov vortices in a spin-singlet superconductor[26]. This is epitomized in superfluid $^3$He with at least seven types of vortices[27]. While such vortices of in $^3$He have been probed spectroscopically via nuclear magnetic resonance[27,28], atomic scale visualization of vortices and their bound states in a charged spin-triplet superfluid has not been possible.

Recently, the heavy-fermion metal UTe$_2$ has been identified as a spin-triplet superconductor based on strong evidence including an abnormally large upper critical field[8,11], direction-dependent Knight shift[29], and magnetic field boosted superconductivity[8,30]. As an orthorhombic crystal (Fig. 1a), UTe$_2$ has a D$_{2h}$ point group symmetry allowing four irreducible representations (irreps), A$_u$, B$_{1u}$, B$_{2u}$, and B$_{3u}$, for odd-parity pairing[31]. Most intriguingly, signatures of time-reversal symmetry breaking (TRSB) were observed from edge-asymmetric tunneling spectra[5], a nonzero polar Kerr effect[4,32], and two superconducting transitions[4] in crystals with $T_c \approx 1.6$ K and residual resistivity ratios (RRRs) around 40. However, the highest quality UTe$_2$ crystals, synthesized via a molten-salt-flux technique[12], display a single superconducting transition and no sign of TRSB[9,10]. Consequently, efforts towards understanding the microscopic details of superconductivity in UTe$_2$ have been hindered by inconsistent experimental results. In this report, we use ultra-clean UTe$_2$ single crystals with $T_c = 2.1$ K and RRR $\approx 680$ (Extended Data Fig. 1) to explore the intrinsic vortex properties of UTe$_2$.

**Vortex visualization via $d^2I/dV^2$ imaging**

UTe$_2$ single crystals have an easy-cleave plane of (011) and a mirror plane perpendicular to the crystallographic **a**-axis (Fig. 1a). A typical STM topographic image (Fig. 1b) reveals atomic chains of Te running along the **a**-axis. We have further defined an in-plane **b**$^*$-axis along the [01-1] direction perpendicular to **a**. To introduce vortices, a magnetic field $B$ is applied perpendicular to the (011) surface, of which the normal direction is denoted as $\hat{\mathbf{n}}_{011}$. Under $B = 4$ T, which is significantly smaller than $H_{c2} \approx 20$ T (ref. 11), the superconducting gap ($|\Delta| \approx 270 - 300$ µeV) far away from vortices (Fig. 1c) remains almost unaltered (or slightly enhanced) from its zero-field value (Extended Data Fig. 2) and is slightly larger than that (~250 µeV)[5,7,31] in UTe$_2$ with $T_c \approx 1.6$ K. Similar to previous reports, the gap is filled with considerable residual DOS that is spatially nonuniform at $B = 0$ T (Extended Data Fig. 2). As a result, although vortices typically have higher zero-bias differential conductance [$g(\mathbf{r}, V) \equiv dI/dV$] (Fig. 1c) due to vortex bound states, conventional vortex visualization via $g(\mathbf{r}, 0 \text{ V})$ and $g(\mathbf{r}, \pm|\Delta|/e \approx \pm 300 \text{ µV})$ imaging is hindered by the heterogeneous in-gap states and yields vortex images with compromised signal-to-noise ratios (Fig. 1d, under 4 T). Nonetheless, the effect of background DOS can be drastically suppressed via $dg/dV(\mathbf{r}, V)$ imaging at $V \approx \pm|\Delta|/2e$ (Fig. 1c). Indeed, a triangular lattice of highly elliptical vortices (indicated by black ellipses) is clearly resolved in the simultaneously acquired $dg/dV(\mathbf{r}, \pm 150 \text{ µV})$ in Fig. 1e with inter-vortex distance $d$, while $dg/dV(\mathbf{r}, 0 \text{ V})$ has minimal contrast, as expected.

**Time-reversal invariant superconductivity under zero magnetic field**

We now turn to the debated question of whether UTe$_2$ has a spontaneous TRSB superconducting order parameter (OP). A TRSB state necessarily requires a linear combination of two D$_{2h}$ order parameters with a $\pi/2$ phase difference, e.g., A$_u \pm$ iB$_{1u}$, yielding two degenerate ground states $\Delta_\pm$ with opposite chiralities under zero magnetic field[33,34,35]. Such degeneracy is lifted under a magnetic field, making the two chiral states physically distinct. For a vortex, four combinations of chirality (C) and vorticity (V) (i.e., defined by the supercurrent winding direction, which is determined by the external magnetic field) are possible, $(C, V) = (++), (+-), (-+)$, or $(--)$, while each time-reversed pair is equivalent [e.g., $(+-)$ and $(-+)$]. Therefore, depending on the combination of vorticity (determined by the field direction) and chirality of the ground state condensate (determined by state preparation), two types of vortices can be expected, each composed of distinct dominant and sub-dominant components with opposite chiralities[33,34,35]. Hence, compared to polar Kerr measurements, wherein the signal might originate from remanent magnetization from trapped vortices upon field-cooling (FC)[32], vortex imaging at the atomic scale constitutes a direct and unambiguous examination of TRS.

Assuming a spontaneous TRSB OP, we first perform FC of UTe$_2$ to 0.3 K under $B = 8.8$ T, preparing the system into a hypothetical $(++)$ state. The magnetic field is decreased to zero, through the Meissner state, inverted to –8.8 T, and cycled back to 8.8 T. At each stage, vortices are imaged in the same field of view (FOV) as shown in Fig. 2a. It is worth noting that if the ground state of a condensate is prepared to have a certain chirality, flipping the direction of the magnetic field below H$_{c,2}$ (i.e., changing vorticity) will not change its chirality as there is no continuous symmetry connecting the states[33]. Therefore, the vortices under positive and negative magnetic fields should correspond to $(++)$ and $(+-)$ configurations with different vortex structures that are discernable even considering finite imaging temperatures[36]. However, the lack of detectable differences between vortices and VLs in Fig. 2a under opposite magnetic fields suggest the absence of spontaneous TRSB. This is further supported by comparing the areal integrated $g(\mathbf{r}, 0\text{ V})$ (Extended Data Fig. 3) between opposite magnetic fields in Fig. 2b that are expected to differ in a chiral superconductor[34,37], yet the differences are smaller than 0.3% for both $|B| = 4$ and 8 T. Next, instead of varying vorticity, we prepared the condensate into two hypothetically different chiralities by FC under $\pm 8.8$ T and imaged the vortices under a series of positive magnetic fields (i.e., same vorticity) in the same FOV. While the vortices should correspond to $(++)$ and $(-+)$ configurations with different vortex structures and consequently VLs, triangular VLs with consistent inter-vortex distances $d$ and vortex structures without statistically significant differences are observed at all magnetic fields for the two cases (Extended Data Fig. 4). Those observations therefore strongly suggest a time-reversal invariant OP in UTe$_2$ under zero magnetic field.

Under a finite magnetic field, on the other hand, multi-component superconductivity is often induced[38,39,40,41,42]. This happens when the crystalline symmetry group of the system is lowered to a subgroup due to the breaking of spin-dependent symmetries, such as mirror reflections and rotations, by the Zeeman splitting. This field-induced symmetry reduction naturally allows the mixing of OPs from different irreps of the original point group. Such a scenario has been discussed in the context of UTe$_2$[38,39]. Furthermore, experimental evidence of continuous $\mathbf{d}$-vector rotation[43,44] and field-trainable Kerr effects[10] in UTe$_2$ are consistent with a field-indued multi-component OP. In our case, the magnetic field is applied along $\hat{\mathbf{n}}_{011}$, which is neither parallel to a rotation axis nor perpendicular to a mirror plane. The resulting Zeeman splitting breaks all the rotation and mirror symmetries in UTe$_2$'s zero-field point group D$_{2h}$, preserving only the inversion symmetry. Because the point group is lowered from D$_{2h}$ to C$_i$, the OP can be a mixture of all odd-parity irreps (A$_u$, B$_{1u}$, B$_{2u}$, and B$_{3u}$), where the specific coefficients depend on energetics. Consequently, with a magnetic field applied along the $\hat{\mathbf{n}}_{011}$ direction, we expect a field-induced multi-component OP in UTe$_2$ with either (1) preserved or (2) broken TRS. Note that our observation of a single type of vortices under different field polarities and cooling history (Fig. 2) is consistent with both possibilities: it does not rule out the possibility of a field-induced chiral OP, where the TRS is explicitly broken by the field. In contrast to a spontaneous TRSB superconductor under zero field, of which the chirality is not tied to the field direction, the chirality of a field-induced chiral OP is dictated to follow the direction of the magnetic field. Therefore, the vorticity and chirality of a superconducting vortex will always have the same sign [i.e., $(C, V) = (++)$ or $(--)$], leading to only one type of vortices that is in agreement with experimental observations (Fig. 2).

## Vortex core bound states

Because the vortices of UTe$_2$ form a triangular lattice, the magnetic flux enclosed by each vortex can be determined by measuring the inter-vortex distance. The result shown in Fig. 2c with both field polarities is in close agreement with single-flux-quantum vortices ($d \approx 1.07\sqrt{\Phi_0/B}$ for triangular lattice, where $\Phi_0$ is flux quantum). In conventional vortices, non-topological Fermionic excitations known as Caroli-de Gennes-Matricon (CdGM) states have a low-energy spectrum of $E_n = \frac{|\Delta|^2}{E_F}\frac{(2n+1)}{2}$, where $n$ is an integer and $E_F$ is the Fermi energy[19]. Figures 3a and b show the normalized differential conductance spectra $g(V)/g(-520 \text{ μV})$ taken at various fields on and off VCs, respectively. A pronounced ZBP with a full width at half maximum (FWHM) of ~250 μV is observed at low fields (red double arrows; we have taken the residual in-gap DOS as a constant background), and gets gradually broadened as the field is increased further. Such FWHM is slightly smaller than those of ZBPs interpreted as MZMs[19,45] in the vortices of FeTe$_{0.55}$Se$_{0.45}$. Surprisingly, the ZBPs in the VCs of UTe$_2$ persist up to 8 T (Extended Data Fig. 5), exceeding the Pauli paramagnetic limit of UTe$_2$ $H_P = 1.86T_c \approx 3.9$ T. Although discrete CdGM states with an energy separation of $\frac{|\Delta|^2}{E_F} \approx 14$ μeV (see Methods) cannot be resolved experimentally, the Zeeman splitting at 8 T ($gu_B B = 0.93$ meV assuming $g = 2$; $u_B$ is Bohr magneton) would greatly exceed the superconducting gap (~0.27 meV). This suggests the observed ZBPs in UTe$_2$ likely correspond to a true zero mode, and thus not CdGM states. Spatially, as shown in the $g(\mathbf{r},V)$ (Fig. 3c) and corresponding $d^2g/dV^2$ (Fig. 3d) spectra taken at B = 4 T across a vortex along the **a**-axis direction, such ZBP extends over 6 nm from the VC without displaying an X-shaped splitting typically observed from CdGM states[20,21] (real space images and more linecuts are shown in Extended Data Fig. 6).

To provide insight on the origin of the robust ZBPs, we consider the two scenarios (1) and (2) separately, where the OP under field is non-chiral and chiral, respectively. In scenario (1), a non-chiral *p*-wave Cooper-pair condensate (with a single- or multi-component OP) is expected to host Majorana Kramers pairs on the boundaries under zero magnetic field[46]. In the presence of a magnetic field, one would naively expect that no MZM pair can survive in a magnetic VC since the TRSB Zeeman splitting would hybridize the pair and push them each to finite energies. However, as a crystalline superconductor, UTe$_2$ possesses a mirror plane (Fig. 1a) perpendicular to the **a**-axis, containing the vortex lines. In fact, crystalline symmetries have been previously proposed to provide topological protection for vortex MZMs in various non-chiral topological superfluid/superconductors[13,21,47]. Here, we theoretically show that integer numbers of MZMs can coexist in a full quantum vortex, where the vortex line can be viewed as an individual 1D system containing additional topological protection from the mirror symmetry.

In the absence of a magnetic field, the superconducting state is described by a mean-field Bogoliubov de Gennes (BdG) Hamiltonian $H_{\text{BdG}} = \sum_{\mathbf{r},\mathbf{r}'} \psi_{\mathbf{r}'}^{\dagger} \begin{pmatrix} H_0 & \Delta \\ \Delta^{\dagger} & -H_0^T \end{pmatrix} \psi_{\mathbf{r}}$ where $\psi_{\mathbf{r}} = (c_{\mathbf{r}\uparrow}, c_{\mathbf{r}\downarrow}, c_{\mathbf{r}\uparrow}^{\dagger}, c_{\mathbf{r}\downarrow}^{\dagger})^T$ is the Nambu spinor, $c_{\mathbf{r}s}^{\dagger}$ creates an electron at position $\mathbf{r}$ with spin $s$, $H_0$ is a spin-degenerate normal state, and $\Delta$ is the non-chiral spin-triplet OP. The superconducting Hamiltonian $H_{\text{BdG}}$ belongs to the topological class DIII[15], which preserves the TRS $\Theta = i\sigma_y \otimes \tau_0 \kappa, \mathbf{r} \to \mathbf{r}$, the particle-hole anti-symmetry $P = \sigma_0 \otimes \tau_x \kappa, \mathbf{r} \to \mathbf{r}$, and an in-plane mirror reflection $M_x = i\sigma_x \otimes \tau_z, (x, y, z) \to (-x, y, z)$, where $\sigma_i$ and $\tau_i$ are Pauli matrices in the spin and Nambu bases and $\kappa$ is the complex conjugate operation. Here, we have renamed $\mathbf{a}$-axis into $x$ and assumed that the OP $\Delta$ is even under reflection $M_x$. When a magnetic field $\mathbf{B} = B\hat{\mathbf{z}}$ is applied, the 1D vortex lines experience a Zeeman splitting $H_z \propto \sigma_z \otimes \tau_z$. As a result, the full Hamiltonian $H'_{\text{BdG}} = H_{\text{BdG}} + H_z$ loses the time-reversal and reflection symmetries since $[\Theta, H_z] \neq 0$ and $[M_x, H_z] \neq 0$. Nonetheless, these vortex lines remain invariant under an effective TRS $\Theta' = M_x \Theta = -i\sigma_z \otimes \tau_z \kappa$ as well as the particle-hole anti-symmetry $P$ since $[\Theta', H'_{\text{BdG}}] = 0$ and $\{P, H'_{\text{BdG}}\} = 0$. Note that this effective TRS $\Theta'$ is fundamentally different from the original $\Theta$ in that instead of $\Theta^2 = -1$, the new symmetry has $(\Theta')^2 = 1$ due to the embedded reflection $M_x$. Therefore, unlike $\Theta$, $\Theta'$ no longer dictates the existence of Majorana Kramers pairs, which is a well-established quantum mechanical consequence[48]. Instead, a vortex line as an independent 1D system belongs to a symmetry-protected topological system in class BDI, which allows any integer numbers of MZMs at the two ends of the vortex line[15], including the case of a single MZM. Although spatially non-splitting ZBPs can also result trivially from anisotropic superconductors[49], considering the persistence of the ZBPs at 8 T and theoretical considerations here together, the observed ZBPs in UTe$_2$ are not inconsistent with MZMs from the topological vortex lines protected by mirror, time-reversal, and particle-hole symmetries in a non-chiral *p*-wave superconductor.

In scenario (2) where a chiral multi-component OP is favored under a magnetic field, the superconducting mean-field Hamiltonian under a magnetic field $H'_{\text{BdG}}$ can break the effective TRS $\Theta'$. This is because in contrast to scenario (1), the $\Theta$-breaking pairing term does not have to preserve $\Theta'$. In the case where $\Theta'$ is broken by the pairing term, the only surviving symmetry in the vortex lines is the particle-hole symmetry P. These vortex lines are thus 1D systems in topological class D, which are topologically equivalent to Kitaev chains and are well expected to host single MZM at the two ends of a vortex line[50], leading to ZBPs surviving up to high magnetic fields.

To first order, the spatial visualization of ZBPs further enables extraction of orientation-dependent superconducting coherence lengths $\xi(\theta)$. The zero-bias conductance image of a single vortex measured at low field (0.5 T) is show in Fig. 3e after two-fold symmetrization [$g_S(\mathbf{r}, 0\text{ V})$] to suppress spatially varying $g(\mathbf{r}, 0\text{ V})$ background under zero-field (Extended Data Fig. 2). $g_S(\mathbf{r}, 0\text{ V})$ is then fitted with an exponential decay $g_S(\mathbf{r}, 0\text{ V}) = De^{-ir/\xi} + F$ for each angle $\theta$, where D and F are constants (see Methods), and the extracted $\xi(\theta)$ is shown in Fig. 3f. Figure 3g shows the results for directions along and perpendicular to the **a**-axis of UTe$_2$, revealing a strong anisotropy with $\xi_a \approx 12$ nm and $\xi_{b^*} \approx 4$ nm. The larger coherence length along the **a**-axis direction contrasts with bulk measurements that determined the upper critical field to be smallest along the **a** axis (i.e., $\xi_a$ is expected to be the smallest)[3,8]. Furthermore, if we calculate the upper critical field along $\hat{\mathbf{n}}_{011}$, we have $H_{c2}^{011} = \frac{\Phi_0}{2\pi \xi_a \xi_{b^*}} = 6.9$ T, which is significantly smaller than that (~20 T) determined from bulk measurements[11]. Such apparent paradoxes could be reconciled if there is more than one superconducting condensate (e.g., existence of multi-band or surface superconductivity), where the one with weaker pairing and a larger coherence length contributes to STM vortex imaging as in the case of MgB$_2$[51]. The weakened surface superconductivity could also be consistent with a competing surface-only charge density wave detected previously[7,52,53]. Moreover, as the magnetic field is increased, the size of vortices decreases following roughly a $1/\sqrt{B}$ trend while preserving a large anisotropy (Extended Data Fig. 7). If UTe$_2$ were a weak-coupling BCS superconductor, this would be consistent with predictions at high fields[54]. Considering the discussions above and similar field-induced vortex-shrinking observed in Cu$_x$Bi$_2$Se$_3$[55], this observation could again be consistent with the presence of distinct and weakened surface superconductivity in UTe$_2$.

**Mirror-asymmetric vortex doublets**

In addition to the high anisotropy, close examination of vortex images reveals that while the mirror symmetry is preserved along the **a**-axis, the mirror symmetry along the **b**$^*$-axis is broken. This is most clearly seen in Figs. 4a,b. The locations of vortices (black ellipse) can be identified in $g(\mathbf{r}, 0\text{ V})$ in Fig. 4a under $B = 4$ T. Simultaneously acquired $dg/dV(\mathbf{r}, 267\ \mu\text{V})$ image in Fig. 4b clearly shows a doublet structure for each vortex indicated by pairs of blue-pink ellipses. Phenomenologically, such a doublet structure undergoes a transition at $|B|\sim 2.5$ T (i.e., blue-pink vs pink-blue; Extended Data Figs. 8, 9) as quantitatively depicted in Fig. 4c. Here, we have plotted the mirror-asymmetry index $\frac{L}{d} - 0.5$ as a function of $B$, where the length $L$ is schematically defined in Fig. 4b. Applying a small in-plane magnetic field in different directions does not alter such doublet structures within measurement precision (Extended Data Fig. 10), suggesting the asymmetry cannot originate from small misalignment between the magnetic field direction and the $\hat{\mathbf{n}}_{011}$ direction of UTe$_2$. While $dg/dV$ imaging provides the first signature of exotic vortex structures, we explore this further by extracting the apparent superconducting gap $\Delta^*(\mathbf{r})$ in Fig. 4d (see Methods) via spectroscopic imaging over multiple vortices. The VCs have nearly zero gap magnitudes with one indicated by a black ellipse. Unprecedent in known superconductors (to our knowledge), although $\Delta^*$ recovers to around the zero-field gap of 270 μeV while moving away from the VC in the $+\mathbf{b}^*$ direction, $\Delta^*$ increases up to 400 μeV in the $-\mathbf{b}^*$ direction and forms a crescent domain of similar size to the VC with a mutual separation of $W = 5 \sim 10$ nm (Fig. 4d and Extended Data Fig. 8). Such crescent-shaped structure is also manifested in $dg/dV(\mathbf{r}, -333\ \mu V)$ in Fig. 4e. To visualize such enhanced apparent gap, differential conductance spectra taken at various locations around a VC (red crosses in Fig. 4d) as well as under zero magnetic field are compared in Fig. 4f, where an enlarged gap with a similarly shaped spectrum is seen in the crescent domain (left of VC). Figures 4g,h display a series of $g(\mathbf{r}, V)$ and $d^2g/dV^2(\mathbf{r}, V)$ spectra taken along a line parallel to the $\mathbf{b}^*$-axis as indicated in Fig. 4a. Clearly, the recovery of $\Delta^*$ on the two sides of a vortex is highly asymmetric, with an overshooting $\Delta^*$ on the left side of each vortex, regardless of field strength and direction (Extended Data Figs. 6, 8, 11, 12).

**Possible origins of asymmetric vortex structures**

The field-dependence of the mirror-asymmetry of vortex structures can shed light on the possible origin of it. As shown in Fig. 4c and Extended Data Figs. 8,9,11, the apparent asymmetry in both $\frac{dg}{dV}$ imaging (represented by $\frac{L}{d} - 0.5$) and the apparent gap $\Delta^*$ remains unchanged (i.e., no flipping) under the switching of magnetic field directions. If magnetic vortex lines were to extend through the UTe$_2$ crystal along $\hat{\mathbf{n}}_{011}$ with preserved mirror-asymmetry, we would expect the asymmetry to be opposite if the direction of $B$ is flipped, contradicting experiments (Extended Data Fig. 13). This suggests instead mechanisms that lock the vortex asymmetry to the lattice asymmetry on the (011) surface. We first consider trivial possibilities. For example, the vortex asymmetry could result from tilted vortex lines considering the strong magnetic anisotropy[2] of UTe$_2$ (Extended Data Fig. 13). However, significant changes of $\Delta^*$ up to 100 μeV (Extended Data Fig. 12) are unlikely in this scenario. Naively, another possibility is a mirror-asymmetric surface Doppler effect[56,57] from a B-field induced anomalous supercurrent flowing on the surface with a velocity $\mathbf{v}_a$ around the $\mathbf{b}^*$-axis (Extended Data Fig. 14). This will result in different net superfluid velocities $v_s$ on the two sides of a vortex and therefore different coherence peak splitting that is twice of the Galilean energy boost $\delta E_{k_F} = \hbar \mathbf{k}_F \cdot \mathbf{v}_s$ (Methods), where $\mathbf{k}_F$ is the Fermi wavevector. Indeed, such considerations are generally true and by assuming an isotropic superconducting gap and circular Fermi surface, one can obtain $v_a \approx 12.6$ m/s (Methods, Extended Data Fig. 14). However, in the unique case of anisotropic UTe$_2$ where the maximum superconducting gap lies along the $\mathbf{b}^*$-axis (Fig. 3f) — perpendicular to the hypothetical anomalous flow along the $\mathbf{a}$-axis— we have $\delta E_{k_F} = 0$ for $\mathbf{k}_F = \pm k_F \hat{\mathbf{b}}^*$. This suggests no splitting of the coherence peak, which reflects the largest gap at $\mathbf{k}_F = \pm k_F \hat{\mathbf{b}}^*$, and thus no change of the apparent gap $\Delta^*$ (Extended Data Fig. 14). Therefore, the vortex asymmetry is unlikely a consequence of the Doppler effect from an anomalous surface supercurrent. It is also unlikely that a weaker surface superconductivity postulated above is responsible for the observed asymmetric vortices with an enhanced $\Delta^*$ and sharper coherence peaks on one side of the VC compared to zero-field measurements (Fig. 4f and Extended Data Figs. 6, 8, 11, 12). Another candidate is the double-core vortices in non-chiral *p*-wave superfluid ³He-B under low temperature and pressure[27,58]. The double-core vortices are each composed of two half-quantum vortices (HQVs) that bound to each other. While HQVs are recently predicted[59] to exist in UTe₂ if assuming a chiral *p*-wave OP, the crescent domain has an enhanced gap $\Delta^*$ as opposed to a suppressed OP, precluding the possibility of the observed doublets being double-core vortices. Recent measurements using superconducting quantum interference devices on UTe₂ also detected no signatures of HQVs[60].

We now discuss our conjectures for the origins of the observed vortex doublet. As mentioned earlier, the applied magnetic field along $\hat{\mathbf{n}}_{011}$ is expected to induce a multi-component OP, which has been known for hosting complex and mirror-asymmetric vortex structures in various multi-component superfluid and superconductors[33,61,62]. This is because regardless of whether the multi-component OP breaks TRS, a subdominant OP can be induced near the VCs of a dominant OP. While the dominant components are generally expected to form vortices with symmetric shapes, depending on the crystalline symmetry group, the induced OP can energetically favor an axis-asymmetric real-space pattern, leading to an asymmetric superconducting regime nearby. In fact, mirror-asymmetric vortex structures of this kind (e.g., triangular or crescent[33,61]) have been predicted in topological superfluid ³He (ref. 61) and unconventional superconductors with different crystalline symmetries, including UPt₃ (ref. 33) and cuprates with an induced *s*-wave component around a *d*-wave core[62]. The observed UTe₂ vortex doublet thus resembles a vortex structure of this kind. Furthermore, the same conjecture of a field-induced multi-component OP, which likely contains the fully gapped irrep $A_u$ due to energetics, can possibly account for field-induced sharper coherence peaks (Fig. 4f, Extended Data Fig. 6).

**Conclusions**

In ultra-pure UTe₂, we uncover anisotropic, time-reversal invariant superconductivity at zero magnetic field, along with persistent VC ZBPs up to 8 T, signatures of surface superconductivity, and mirror-asymmetric vortex doublets—each comprising domains with suppressed and enhanced superconducting gaps. Theoretical analysis suggests the robust ZBP to be consistent with symmetry-protected MZMs. We propose that a field-induced multi-component OP, arising from the reduced point group symmetry, can potentially lead to the emergence of mirror-asymmetric vortices by condensing a subdominant OP. The interplay of anisotropic superconductivity, spin-triplet pairing, spin-orbit coupling, and **d** vector rotations under a magnetic field implies rich physics to be explored but also challenges for theoretical modeling. The exact forms of OPs and their vortex solutions under magnetic fields deserve future exploration.

**Methods**

Crystal growth and transport measurements

Single crystals of $UTe_2$ were grown through a molten salt technique using an equimolar mixture of sodium chloride (NaCl) and potassium chloride (KCl) as reported previously (ref. 12). The crystallographic structure of our crystals was verified at room temperature by a Bruker D8 Venture single-crystal x-ray diffractometer equipped with Mo K-$\alpha$ radiation. To ensure that the samples only show a single superconducting transition temperature, specific heat measurements were performed using a Quantum Design calorimeter that utilizes a quasi-adiabatic thermal relaxation technique. Electrical resistivity was measured with an alternating current resistance bridge using a standard four-point technique with current along the [100] direction. The RRR is extracted by fitting the temperature-dependent resistivity $\rho(T)$ with $\rho = AT^2 + \rho_0$ in the low-temperature region before the superconducting transition (Extended Data Fig. 1). Then RRR is calculated by RRR = $\rho(300 \text{ K})/\rho_0$.

Scanning tunneling microscopy and spectroscopy

STM experiments are performed on a Unisoku USM1300J systems at a base temperature of 0.3 K. The UTe2 crystals are cleaved at a temperature of ~ 10 K under ultrahigh vacuum (~1 × $10^{-10}$ Torr) before loaded into the STM head. Both mechanically cut Nb and PtIr tips are used for measurements after thorough outgassing via heating in vacuum. The zero-field differential conductance spectrum shown in Fig. 4f and the map shown in Extended Data Fig. 2 are measured with PtIr tips, while all other data are acquired using Nb tips. While the Nb tips are superconductive under zero magnetic field and allow us to extract the $UTe_2$ gap via multiple Andreev reflections (Extended Data Fig. 2), the Nb tips become normal and metallic under all magnetic fields (> 0.3 T) reported in this study such that no deconvolution of spectra is necessary. Field-cooling is performed by applying a magnetic field at a temperature above $T_c$ followed by cooling down to the base temperature of 0.3 K. STM data are acquired using SPECS Nanonis electronics. Spectroscopic measurements ($I - V$, $g - V$, and $dg/dV$) are performed using a built-in lock-in amplifier in Nanonis with a modulation of 100 μV, while $dg/dV$ spectra can also be acquired via numerical derivative of $g - V$ spectra. $d^2g/dV^2$ spectra are all obtained via numerical derivative of $g - V$ spectra. A typical lock-in modulation frequency is 983.7 Hz. MATLAB and Gwyddion software are used for data processing.

Coherence length extraction

The $g(\mathbf{r}, 0 \text{ V})$ images are first smoothened by a Gaussian filter of 0.5 pixel to reduce random noise, and then symmetrized in a twofold fashion along the **a** and $\mathbf{b}^*$ directions to generate $g_S(\mathbf{r}, 0 \text{ V})$. Strictly speaking, because of the mirror-asymmetric vortices, symmetrization along the $\mathbf{b}^*$ direction is not justified. However, to first order, this approach helps suppress spatially varying in-gap DOS existing at zero field, providing a good estimate of the coherence lengths. Then, for each vortex, the $g_S(\mathbf{r}, 0\,\text{V})$ image is fitted with a function $g_S(\mathbf{r}, 0\,\text{V}) = \text{D}e^{-r/\xi} + \text{F}$ along different directions, where the origin of $\mathbf{r}$ is defined to be the center of each VC. The constant F represents and is determined by the zero-energy DOS far from each vortex, and the constant D is a scaling factor.

Superconducting gap extraction

To precisely extract the apparent superconducting gaps $\Delta^*$, defined as half of the voltage separation between two coherence peaks in differential conductance spectra, we use a curvature-based analysis method[63] that have shown effectiveness in peak extractions in STM[64]. Each point spectrum $g(V)$ is first interpolated using a cubic spline fit. Then, a local curvature of each spectrum is calculated as

$$C(V) = -\frac{d^2 g/V^2}{[C_0 + (dg/dV)^2]^{3/2}} \qquad (2)$$

where $C_0$ is a constant. In this way, the peak positions of $g$ can be extracted accurately from the local maxima of $C$. ZBPs detected within a voltage range of $-150\,\mu\text{V}$ to $150\,\mu\text{V}$ indicate the location of VCs. The validity of detected peaks is verified manually. Once the ZBP locations are determined, the gap is set to zero at those locations. To extract $\Delta^*$, locations of coherence peaks are detected in the energy ranges of $(-450\,\mu\text{V}, -200\,\mu\text{V})$ and $(200\,\mu\text{V}$ to $450\,\mu\text{V})$. The extracted gap maps are manually verified.

Estimation of energy separation of CdGM states

Discrete CdGM states[19,65] have energy separations of $\delta E_n = \frac{|\Delta|^2}{E_F}$. Because $E_F = \frac{\hbar k_F v_F}{2}$ and $\xi = \frac{\hbar v_F}{\pi |\Delta|}$, we have $\frac{|\Delta|}{E_F} = \frac{2}{\xi \pi k_F}$ and thus $\delta E_n = \frac{2|\Delta|}{\xi \pi k_F}$. For UTe$_2$, we can take $k_F \sim 3/\text{nm}$ (ref. 66), $|\Delta| = 0.27\,\text{meV}$, and $\xi = 4\,\text{nm}$, and obtain $\frac{|\Delta|}{E_F} \approx 0.05$. Since $\frac{T}{T_c} = 0.14$ for $T = 0.3\,\text{K}$, the quantum limit situation ($\frac{T}{T_c} \ll \frac{\Delta}{E_F}$) is not satisfied[65] and $\delta E_n = \frac{|\Delta|^2}{E_F} \approx 14\,\mu\text{eV}$ is beyond the energy resolution at 0.3 K, where $k_B T = 26\,\mu\text{eV}$.

DOS simulation considering the Doppler effect

The calculation and extraction of superfluid flow velocity by measuring quasiparticle DOS is detailed in Ref. 56. Briefly, assuming a circular Fermi surface, a Galilean energy boost $\delta E_k \equiv \hbar \mathbf{k} \cdot \mathbf{v}_s(\mathbf{r})$ will alter the spectrum of a quasiparticle $|\mathbf{k}\rangle$ in a flowing superfluid: $E_k = \pm\sqrt{\varepsilon_k^2 + |\Delta|^2} + \hbar \mathbf{k} \cdot \mathbf{v}_s$, where $\varepsilon_k$ is the normal-state band structure and $\mathbf{v}_s$ is the superfluid velocity. Using the Dynes formula, the quasiparticle DOS with a finite superfluid flow can be written as

$$N(E) = \text{sign}(E) \int_0^{2\pi} Re\left(\frac{E - \delta E_{\boldsymbol{k}} - i\gamma}{\sqrt{(E - \delta E_{\boldsymbol{k}} - i\gamma)^2 - |\Delta(\theta)|^2}}\right) d\theta \qquad (3)$$

The consequence of a finite $\delta E_{\boldsymbol{k}_\text{F}}$ for $\boldsymbol{k}_\text{F}$ along the direction of the largest superconducting gap is that the superconducting coherence peak exhibits a symmetric splitting of $2|\delta E_{\boldsymbol{k}_\text{F}}|$, leading to a decreased $\Delta^*$ by $|\delta E_{\boldsymbol{k}_\text{F}}|$ and reduced coherence peak heights as observed previously[56,57] and shown in Extended Fig. 14.

**Acknowledgements**
The authors thank M. Eskildsen, J. Sauls, and N. Gluscevich for valuable discussions. N.S., M.T, J.M., F.C. and X.L. acknowledge support from the U.S. Department of Energy (DOE), Office of Science, Office of Basic Energy Sciences under Award Numbers DE-SC0025021 (2nd derivative vortex imaging) and DE-SC0024291 (general STM operation), and the Stavropoulos Center for Complex Quantum Matter at the University of Notre Dame. Y.-T.H. acknowledges support from National Science Foundation Grant No. DMR-2238748. N.S. and M.T. acknowledge support from the Notre Dame Materials Science and Engineering Fellowship. Work at Los Alamos was supported by the U.S. Department of Energy, Office of Science, National Quantum Information Science Research Centers. M.M.B. acknowledges support from the Los Alamos Laboratory Directed Research and Development program.

**Author contributions**
X.L. conceived the project. N.S., M.T., and J.M. performed the measurements. M.M. B, S.M.T., and P.F.S.R. synthesized the crystals and performed electrical resistivity measurements. Y.-T.H. performed theoretical analysis. N.S., M. T., J.M. and F.C. performed data analysis. X.L. and Y.-T.H. wrote the manuscript with input from all authors. All authors contributed to data interpretation.

**Data availability**
All data needed to evaluate the conclusions in the paper are present in the paper and/or the Extended Data. Additional data related to this paper may be requested from the authors.

**Competing financial interests**
The authors declare no competing financial interests.

# Figures

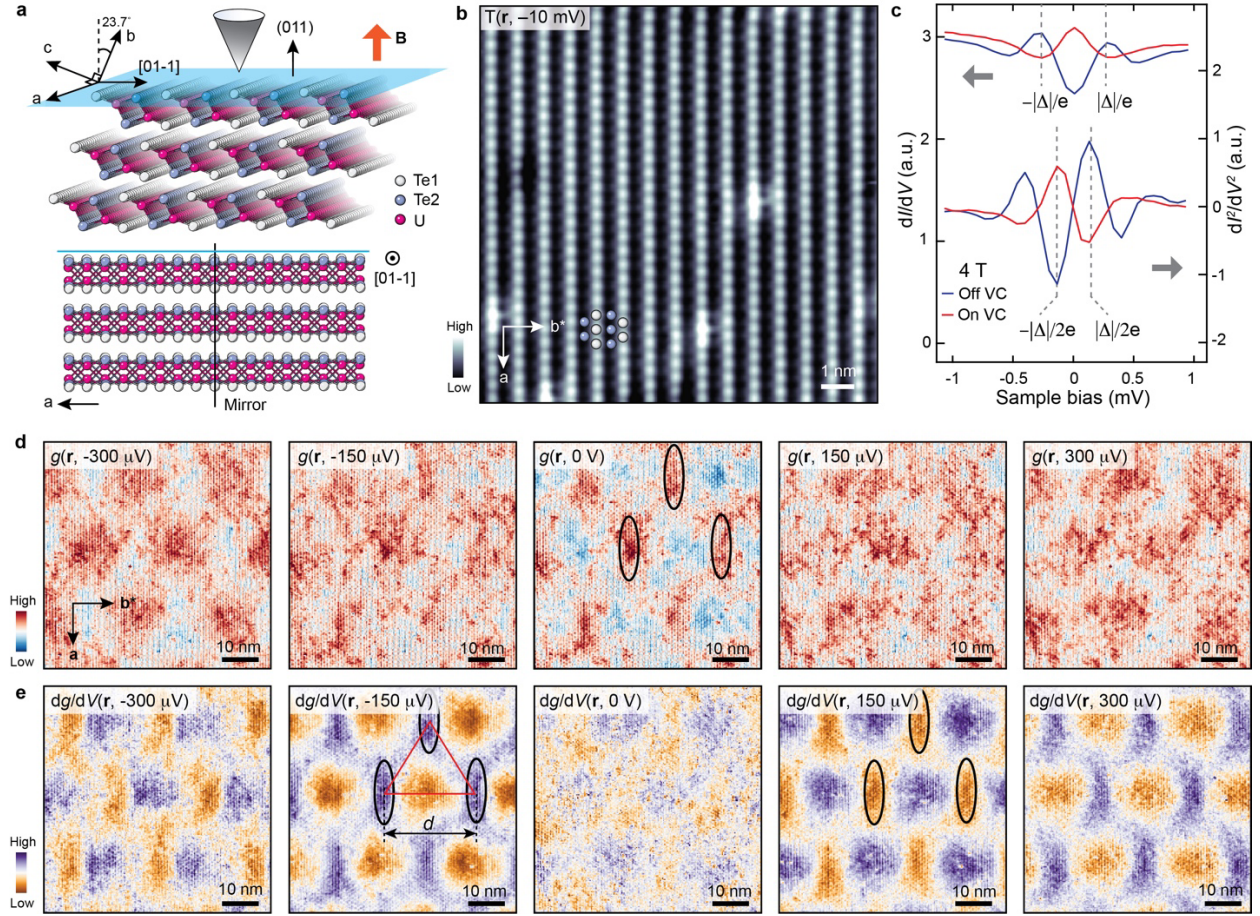

**Figure 1. Visualizing superconducting vortices of UTe$_2$.**
**a**, Top: schematic of measurement setup. STM images the (011) surface of UTe$_2$ with the magnetic field applied in the normal direction of (011) surface (defined as $\hat{\mathbf{n}}_{011}$ direction), which is 23.7° rotated from the crystallographic b axis in the b-c plane. Bottom: a side view of the UTe$_2$ crystal with the magnetic field direction pointing vertically up. A mirror plane containing the vortex lines is indicated. **b**, A typical topographic image of UTe$_2$ (011) surface with overlaid Te1 and Te2 atoms (setpoint: $V_S = -10$ mV, $I_0 = 1$ nA). **c**, Differential conductance ($g$) and $dg/dV$ spectra taken on and off VCs of UTe$_2$ under $B = 4$ T. Maximal contrast of vortices are expected at energies around $\pm|\Delta|/2e$. **d**, $g(\mathbf{r}, V)$ and **e**, $dg/dV(\mathbf{r}, V)$ images acquired at $B = 4$ T. The triangular lattice of vortices are indicated by black ellipses with inter-vortex distance $d$. The vortices are best visualized in $dg/dV(\mathbf{r}, V \approx \pm|\Delta|/2e \approx \pm 150$ μV).

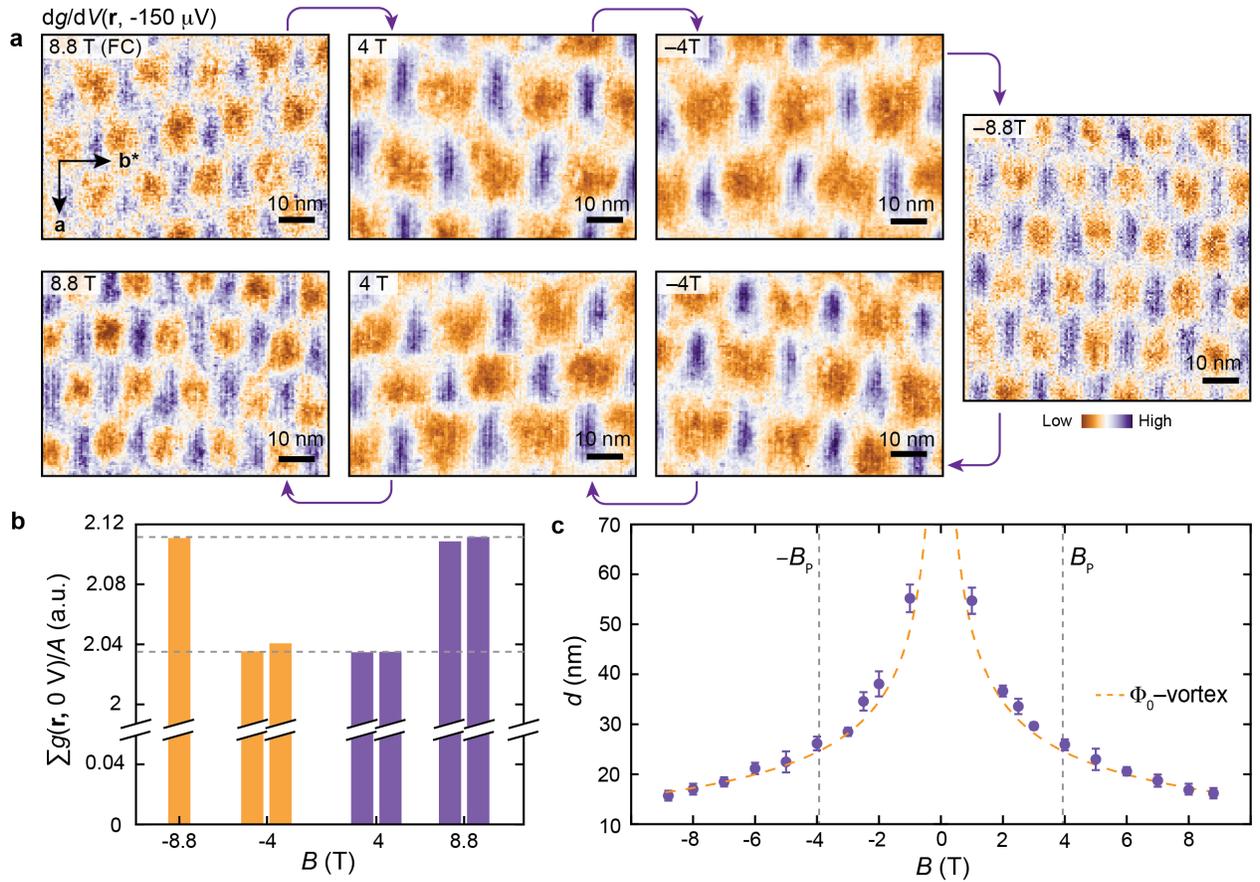

**Figure 2. Time-reversal invariant superconductivity**
**a**, A series of $dg/dV(\mathbf{r}, -150\,\mu\text{V})$ images of vortices at different fields under a single field-cooling (FC) with $B = 8.8$ T. No discernable differences of the vortex structure can be observed. The slight tilts of the VLs are likely due to slow vortex creeping. **b**, Areal integrated zero-energy DOS at different fields, showing differences < 0.3% between different field polarities. **c**, Extracted inter-vortex distance as a function of $B$, showing agreement with single-flux-quantum ($\Phi_0$) vortices. The vertical lines indicate the Pauli paramagnetic limit $B_\text{P} \approx 3.9$ T.

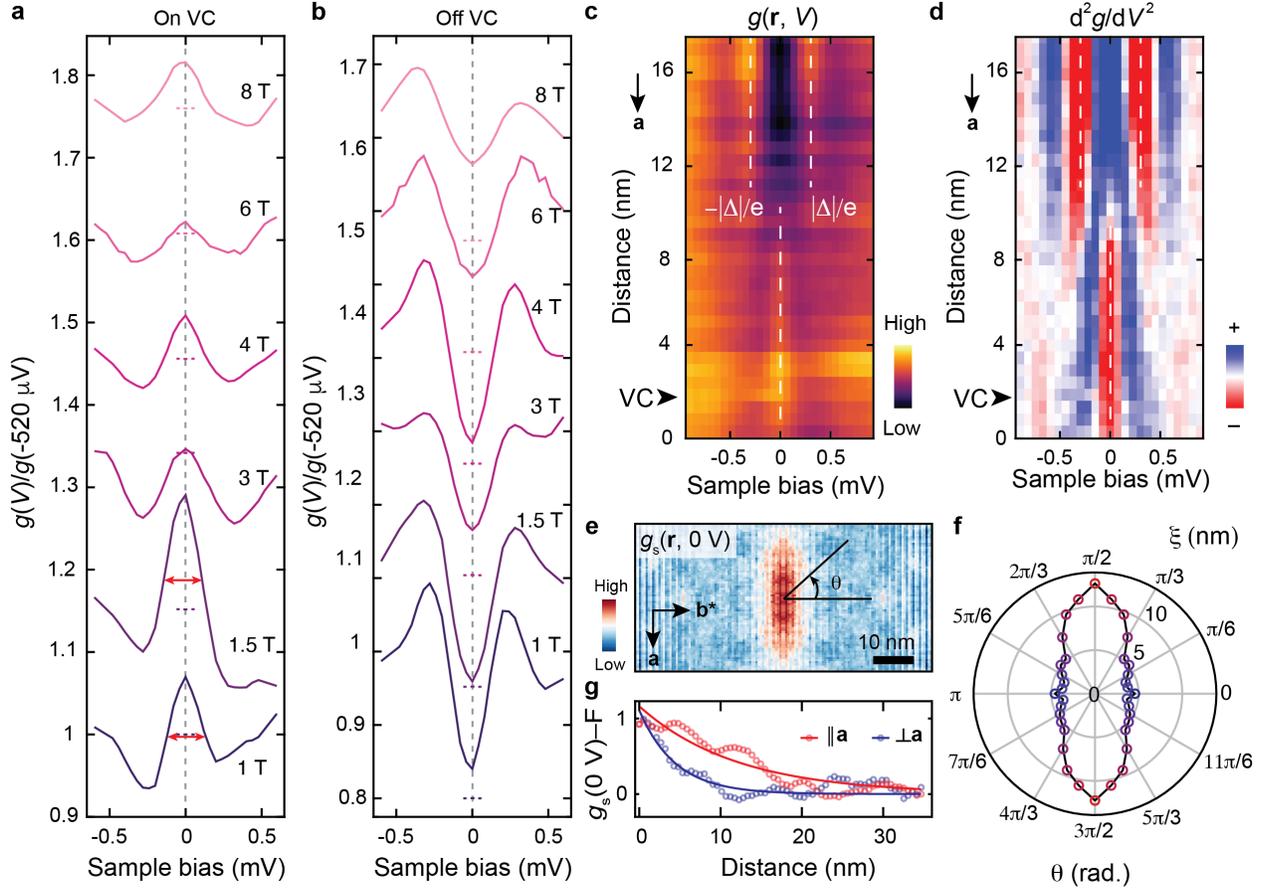

**Figure 3. Persistent zero-bias peak and coherence lengths extraction**
**a**, Normalized differential conductance spectra $g(V)/g(-520\,\mu V)$ measured at the VCs under different magnetic fields showing prominent ZBPs. The dashed lines indicate a level of $g(V)/g(-520\,\mu V) = 1$ for each spectrum. The red double arrows correspond to $250\,\mu V$, indicating the FWHMs of ZBPs at low fields. **b**, Normalized differential conductance spectra $g(V)/g(-520\,\mu V)$ measured far from vortices under different magnetic fields showing almost unaffected superconductivity gap size and gap filling at higher fields. The dashed lines indicate a level of $g(V)/g(-520\,\mu V) = 0.8$ for each spectrum. **c**, A series of $g(V)$ and **d**, $d^2g/dV^2$ spectra taken across a vortex at B = 4 T along the **a**-axis direction (indicated by an arrow) showing a non-splitting ZBP extending over 6 nm in one direction from the core (indicated by an arrow head). **e**, Symmetrized zero-bias conductance image $g_S(\mathbf{r}, 0\,V)$ at B = 0.5 T of a single vortex. The angle θ is defined with respect to the $b^*$ direction. **f**, Extracted angle-dependent coherence length. **g**, $g_S(\mathbf{r}, 0\,V)$ and fittings for directions along and perpendicular to the **a**-axis of UTe$_2$, revealing anisotropic superconductivity with $\xi_a \approx 12$ nm and $\xi_{b^*} \approx 4$ nm.

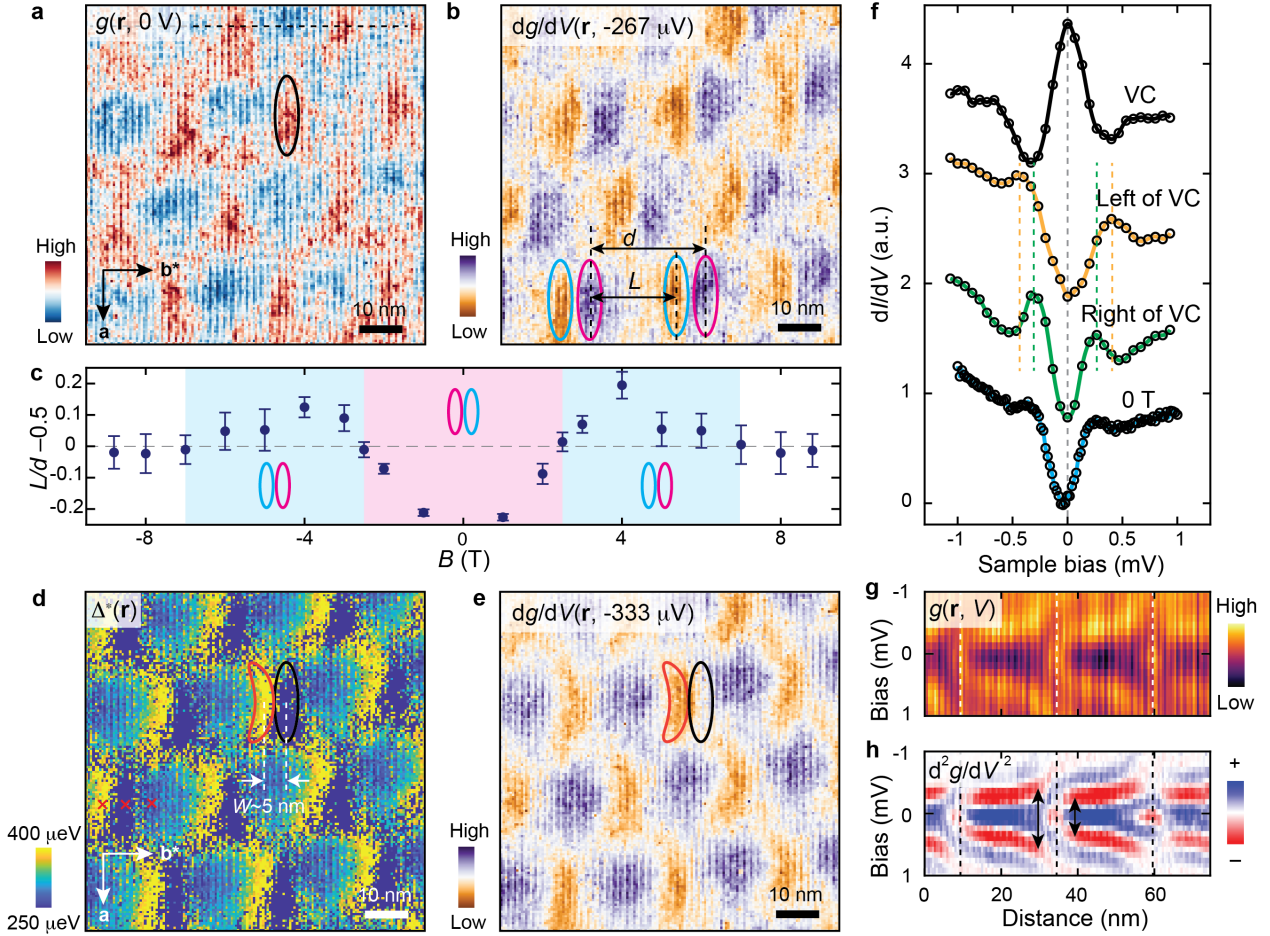

**Figure 4. Mirror-asymmetric vortex doublet**

**a**, Zero-bias conductance image of vortices under B = 4 T. The black ellipse indicates the location of a VC. **b**, Simultaneously acquired $dg/dV(\mathbf{r}, -267\ \mu V)$ image showing each vortex forming a doublet structure as indicated by the blue-pink pairs. Note that the pink ellipses do not necessarily correspond to the locations of VCs (black ellipse). **c**, A transition of the doublet structures as indicated by the blue-pink and pink-blue ellipses is observed symmetrically as a function of magnetic field. This is quantitatively demonstrated using an index defined as $\frac{L}{d} - 0.5$. **d**, Extracted apparent superconducting gap $\Delta^*(\mathbf{r})$ showing a domain of enhanced gap (red crescent) on the left side of each VC (black ellipse), which can also be observed in **e**, simultaneously acquired $dg/dV(\mathbf{r}, -333\ \mu V)$. The separation between the ellipse and the VC is around $\xi_{b^*}$. **f**, $g(V)$ spectra at locations indicated by the red crosses in (d) and at zero magnetic field, demonstrating an enhanced gap up to $\sim 400\ \mu eV$ in the crescent domain. **g**, A series of $g(V)$ and **h**, $d^2g/dV^2$ spectra taken along the dashed line in (a) showing enhanced gaps on the left side of each VC as indicated by the double arrows.